\documentclass{article}

\usepackage{arxiv}
\usepackage{cite}
\usepackage{svg}
\usepackage{mwe}
\usepackage{appendix}

\usepackage[utf8]{inputenc} 
\usepackage[T1]{fontenc}    
\usepackage{hyperref}       
\usepackage{url}            
\usepackage{booktabs}       
\usepackage{amsfonts}       
\usepackage{nicefrac}       
\usepackage{microtype}      
\usepackage{multirow}

\usepackage{array}
\usepackage{authblk}

\title{Globalization emergence in the European Patent Office (EPO) patent network}

\author[1]{Maria Tsouchnika}
\author[2]{Alex Smolyak}
\author[1]{Panos Argyrakis}
\author[2]{Shlomo Havlin}

\affil[1]{Department of Physics, University of Thessaloniki, and Center of Complex Systems, 54124 Thessaloniki, Greece}
\affil[2]{Department of Physics, Bar-Ilan University, 52900 Ramat-Gan, Israel}

\begin{document}
\maketitle

\begin{abstract}
We studied the evolution of the European Patent Office (EPO) patents applicants’ collaborations network, within a 35 years span of data (1978-2013). Focusing on the Giant Component (GC) formation process over many time-windows, distributed throughout the data timeline, we found that the features governing this phenomenon are indicative of emerging globalization in the applicants’ collaborations. The timeline appears to be divided into three regimes, corresponding to three states of the network's evolution. In the early years state, the GC takes long to form and the instant of its creation is easily pinpointed, while it features geographically segregated groups of applicants with technologically similar activities. In contrast, in the late years state, the GC forms quickly, the exact point of its creation is harder to spot, the applicants' activities are more disparate technologically, while their inter-regional collaborations are significantly increased. The middle years are an intermediate state between the two extreme of early and late years. Moreover, we concluded that the critical patents, which induce the GC’s formation, are typically introduced by large-sized applicants and also that top patent-producing applicants are likely to submit critical patents, albeit at a lower rate than their overall patent submission. Lastly, we uncovered the crucial role that Japan plays in the network’s coherence, through its prominent participation in the GC and the critical patents.
\end{abstract}


\section{Introduction}\label{sec:intro}
Innovation is widely regarded as a major contributing factor to economy growth and job creation. A plethora of theoretical and empirical studies support the notion that innovation, in many ways, has a positive effect on productivity and growth~\cite{solow1957technical,griliches1958research,mansfield1961technical,mansfield1962entry,griliches1964research,mansfield1965rates,griliches1980returns,griliches1983comparing,griliches1985productivity,NAP612,romer1986increasing,geroski1989entry,caballero1993high,geroski1993profitability,klette1996r,crepon1998research,harhoff1998r,klette2000accumulation,klomp1999importance,loof2002knowledge,cainelli2004impact,parisi2006productivity,hall2009innovation,cassiman2010innovation,hall2011innovation,mohnen2013innovation,medda2014technological,raymond2015dynamic}. Consequently, fostering innovation is important in improving the living standards and welfare of people and nations.

However, innovation alone does not suffice for promoting well-being; adequate diffusion and adoption of the corresponding knowledge is also required~\cite{mansfield1961technical,damijan2015learning,hall2003adoption}. Diffusion of the innovation’s outcome is a prerequisite for boosting productivity and, consequently, growth~\cite{gurbiel2002impact,surinach2011extension,moreno2014innovation}. An effective way to achieve diffusion of knowledge is through the collaboration between all potential carriers of innovation, i.e. inventors, firms, research institutes, etc. Becheikh et al.~\cite{becheikh2006lessons}, performed a systematic review of the relevant literature, from 1993 to 2003, and concluded that networking is a remarkably good determinant of innovation. This conclusion was based on the fact that the majority of the studies examined, found networking to have a beneficial effect on innovation. The rest of the studies reviewed by Becheikh et al. found the effect to be insignificant, while none revealed a notable negative effect. Furthermore, knowledge flow is often trapped within regional and firm boundaries. Promoting collaborative ties between regions and firms has been shown to significantly increase the chance of knowledge escaping these boundaries~\cite{singh2005collaborative}. Moreover, there is a growing need of forming collaborative ties between the carriers of innovation, which have become more collective and distributed over the past decades~\cite{teece1992competition,tether2002co}. Plausible motives driving this behaviour include pooling resources, tackling difficulties and reducing the risks related to the innovation process, as well as keeping up with the fast pace of technological change and increasing competitiveness~\cite{tether2002co,brown1995product,stock2002firm}.

Apart from networking and forming collaborative ties, another feature which effect on innovation output has been widely investigated is firm size. While in 1934, Schumpeter~\cite{schumpeter1934theory} suggested that small size favours innovation output, he contradicted it eight years later~\cite{schumpeter1942socialism}. This triggered a major debate that led to a series of theoretical and empirical studies, which reach far into the recent years. Indicatively, Tether in~\cite{tether1998small} questions the validity of a belief that small-sized firms are more efficient innovators than large ones~\cite{acs1990innovation}, which emerged in the early 90’s. This belief was prompt by a group of empirical studies~\cite{acs1990innovation,pavitt1987size,kleinknecht1993collecting,santarelli1996analyzing,cogan1993irish}, which all find that small-sized firms produce a higher number of innovations per thousand employees, than large ones do. Moreover, in~\cite{becheikh2006lessons}, while the majority of the relevant studies reviewed are suggestive of a positive correlation between size and innovation output, there is also a number of studies indicative of a negative, negligent, or even a complex relationship.

Overall, the subject of size vs. innovation output is regarded rather inconclusive, and the relationship between these two features is not considered straightforward~\cite{rogers2004networks,revilla2012relation}, but complex, in spite of the fact that most studies reveal a positive correlation~\cite{becheikh2006lessons}. Therefore, many studies tend to examine a more specific aspect of the subject, as the effect of the size on innovation, with respect to networking~\cite{tether2002co,rogers2004networks}, or with respect to the technological regimes~\cite{revilla2012relation}. Another aspect that is particularly interesting is the effect of size on how radical the produced innovations are. In~\cite{minguela2014cooperation}, it is found that small-sized firms commonly engage in more radical innovations, while large-sized in more incremental ones. This result is attributed to the fact that small-sized firms are considered more flexible and adaptable to technological changes than the more rigid large-sized ones. Similarly, Stock et al.~\cite{stock2002firm} used data from the computer modem industry and inferred that small-sized, adjustable firms exhibit higher rate of technological change in the performance of their products.

All of the above motivated us to analyze the patents data from a network perspective, highlighting aspects that promote - or inhibit - collaboration. Patents data, although not short of weaknesses, are recognized as a useful means of determining technological advance and innovation~\cite{loof2002knowledge,pavitt1985patent,griliches1998patent} and have thus been used in many pertinent studies~\cite{caballero1993high,crepon1998research,hall2011innovation,pilkington2004technology,fleming2007small,schilling2007interfirm,ma2008patent,zhang2017network}. A patent serves as a reflection of the underlying innovative procedures that led to the corresponding invention. Furthermore, patent data are stored in readily available and up to date databases, comprising information about the inventions, such as their date and type, and about the people, institutes and firms involved, as well as their collaborations and interactions. Therefore, by studying the network of collaborations formed by all those involved in the production of a patent, we implicitly investigate aspects of the relationship between these collaborative activities and innovation. The nodes of the network are the patent applicants, representing the firms/inventors working on a project that led to a patent application, filed at the European Patent Office (EPO). A link between a pair of nodes (applicants) is drawn, when the applicants have at least one joint patent application. The network of patent collaborations studied is a social network and, more specifically, an affiliation network~\cite{newman2001structure,newman2002random,barabasi2002evolution,albert2002statistical}.

Clearly, having a network with many small isolated components does not promote collaboration and the diffusion of knowledge as much as a network having a giant component (GC) would\footnote{The GC of a network is a connected component which size is proportional to the number of the network's nodes~\cite{newman2001structure,newman2002random,barabasi2002evolution,albert2002statistical}.}. Forging co-operative arrangements for innovation is found to correlate positively with higher levels of innovation~\cite{tether2002co}. There is also evidence implying that a network with short path lengths and increased aggregation of isolated components into bigger ones, has a positive effect on future patent production and therefore on innovation~\cite{fleming2007small,schilling2007interfirm}. Moreover, Bettencourt et al.~\cite{bettencourt2009scientific} introduce and explore the very interesting idea that critical (red-bond) patents - the addition of which results in the formation of the network’s GC, at the percolation threshold - represent a highly innovative moment in the course of scientific events.

In view of the above, the GC of the patents collaboration network is an excellent standpoint from which to study the interplay between collaboration and innovation, by examining questions, such as: \textbf{\textit{Is there a GC in this network? If so, how long it takes it to form? Could we identify some of the features that influence its formation? Do these features change over time and, if so, how?}} And, last but not least, \textbf{\textit{are the major patent contributors (large-sized applicants) more likely than the ones with medium to small contributions (small/medium-sized applicants) to produce a critical patent?}} Probing into the latter question would be an indirect contribution to the size vs. innovation output problem, assuming that the larger the size of an applicant the more patents is capable of producing~\cite{tether2002co}. Overall, investigating the above questions would be one more step towards unraveling the complex role that social collaboration networks play in promoting economic growth and innovation.

Our results are indicative of a three-part division of the available timeline regarding specific characteristics of the GC formation and of the two major groups of applicants (largest and second largest connected components of the growing network) immediately before the percolation threshold. The system appears to undergo a shift in its state as it advances through time. It appears to move from a state of slow, clear-cut percolation transitions, marked by technological similarity and geographical confinement of the two groups, to a state featuring faster, although more incremental, transitions that hallmark technological complementarity and geographical interplay, having passed through a middle, transitive state. Furthermore, our analysis suggests that top patent-producing applicants are likely to introduce critical patents, although not as likely as would be expected according to their overall patent output. Also the vast majority of red-bond applicants are found to be large-sized firms. Lastly, our findings highlight the key-role played by Japan to the EPO patent network and predominantly to its GC.

\section{Method of calculation - Results}
\label{sec:results}

\subsection{Basic analysis of the static network}
\begin{figure}[ht]
    \centering
    \includegraphics[width=1.0\textwidth]{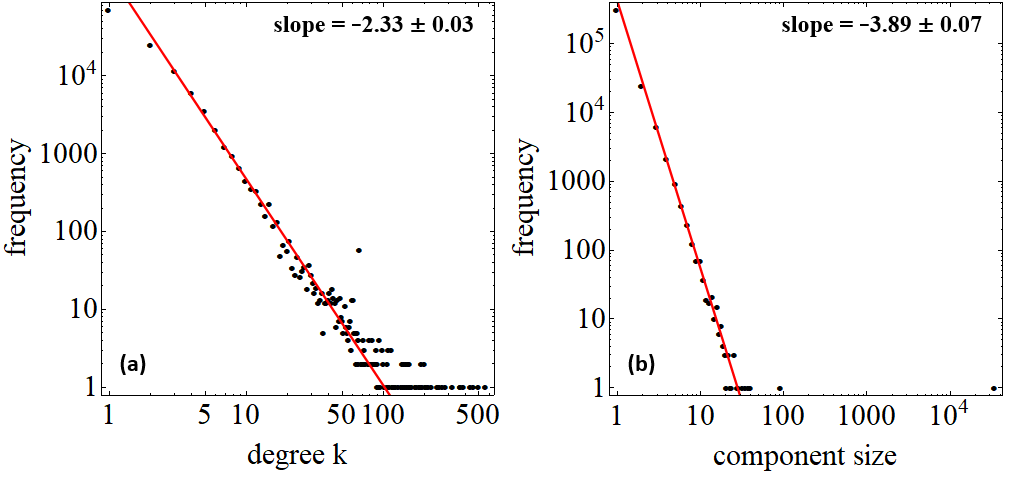}
    \caption[Damage report!]{(a) Degree and (b) component size distributions of the aggregated, static network. \label{fig:katanomes}}
\end{figure}

One of the hallmarks of the patents data\footnote{For details about the data used, see~\ref{sec:epoData}.} is the fact that the vast majority of the patents are filed by individual applicants. Out of the 2,502,311 patents included in the 35 years of data, just 6.2\% (154,474 patents) result from collaborative activities. A basic analysis was performed on the extracted aggregated, static network. It was found to have 429,359 nodes (applicants), consisting of 306,238 isolated nodes (components of size 1) and 123,121 nodes that are connected by 151,474 links (patents). The degree distribution, with a slope of -2.33 and the component size distribution, with a slope of -3.89 are depicted in Fig.~\ref{fig:katanomes}. The maximum and average degree are 852 and 4.05, respectively, whereas the average path length and the network diameter are 5.53 and 27. The largest connected component (LCC) is made up of 34,214 (28\%) nodes and 69,246 links. The second largest component (SLCC) consists of a mere 93 applicants. The isolated applicants of the patents network were omitted in all parts of the subsequent analysis for which they were impertinent. 

\subsection{The emergence of the Giant Component (GC)}\label{subsec:GCemergence}

\begin{figure}
    \centering
    \begin{minipage}{0.49\textwidth}
        \centering
        \includegraphics[width=0.95\textwidth]{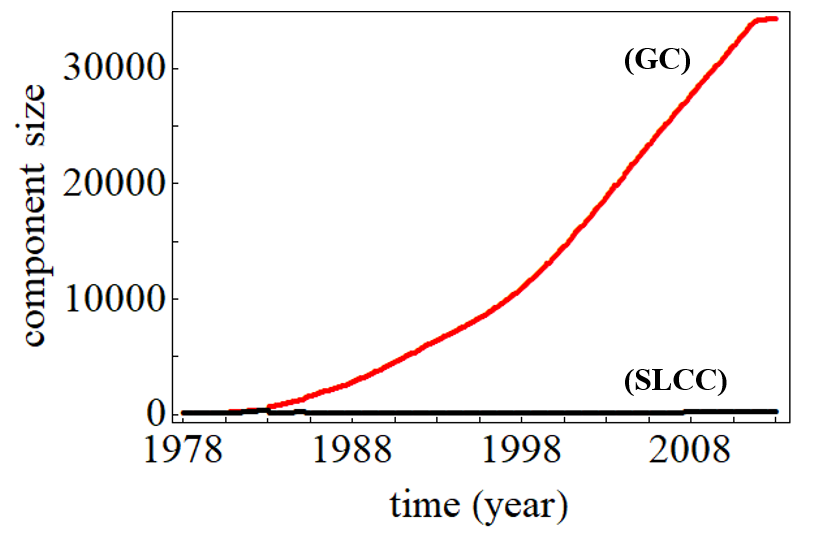} 
        \caption[Damage report!]{Size of a) the largest (red) and b) the second largest (black) connected components of the growing network, throughout the available timeline (1978 – 2013).\label{fig:LCgrowth}}
    \end{minipage}\hfill
    \begin{minipage}{0.49\textwidth}
        \centering
        \includegraphics[width=0.9\textwidth]{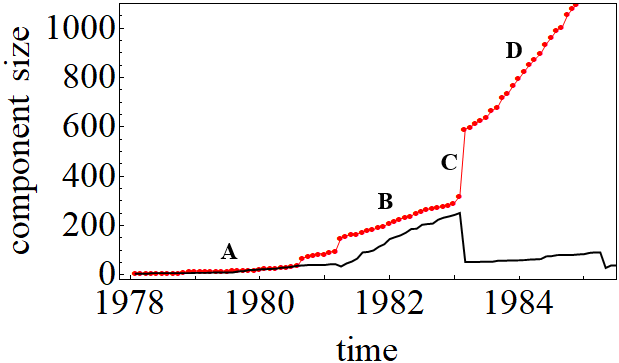} 
        \caption[Damage report!]{The emergence of the giant component. Starting date of the system's temporal reconstruction: 1978, largest component: red circles, second largest component: black line. Phase A: Small-sized, indistinguishable components, phase B: the two major components become discernible, phase C: the percolation threshold, the GC forms, phase D: the GC outgrows all other components.\label{fig:LCgrowthDetail}}
    \end{minipage}
\end{figure}

The massive difference between the LCC and the SLCC hints that the LCC is also the Giant Component (GC) of the network. The GC of a network is a connected component which size is proportional to the number of the network's nodes, in other words \textit{the GC grows with the network}~\cite{newman2001structure,newman2002random,barabasi2002evolution,albert2002statistical}. Fig.~\ref{fig:LCgrowth} depicts the sizes of the largest and second largest connected components of the growing network, over the 35 years of the data. To obtain this result, we replicated the network’s growth, i.e. we progressively added the patents to the system, in a non-descending chronological order. Once a new patent, with at least two applicants\footnote{All one-applicant patents were previously extracted from the data set, as these represent island nodes or self-loops that do not add to any component of the growing network.} is submitted, it contributes to the network’s growth by either adding new nodes (applicants) to existing components, forming new components, or merging existing components into bigger ones. Fig.~\ref{fig:LCgrowth} confirms that the LCC is in fact the GC of the network, as it clearly grows with the network.

The growth process was found to be consistent with that reported in similar previous studies~\cite{newman2001structure,bettencourt2009scientific,liu2015evolutionary,liu2015structure}. Early on, a collection of small-sized components are formed. During this phase all components are indistinguishable with respect to their size. A second phase follows that leads up to the percolation threshold, during which one can discern two major components of comparable size, growing at approximately the same rate.

At the percolation threshold, a single patent that enters the system introduces a critical link that joins these two components into one; the growing GC~\cite{newman2001structure,newman2002random,barabasi2002evolution,albert2002statistical}. From this point onward, the vast majority of new patents join new or existing nodes/components to the GC. Consequently, the evolving GC rapidly outgrows any other component in the system and a vast gap between its size and that of the Second Largest Connected Component (SLCC) is quickly forged, as seen in Fig.~\ref{fig:LCgrowth}. Fig.~\ref{fig:LCgrowthDetail} highlights the growth process outlined above, depicting the emergence of the GC, up until approximately two years after the percolation threshold. It appears that \textit{it takes approximately five years for the two major independent groups of applicants to connect and form the GC}.

What are the implications of the percolation incidence during the growth process? In a \textit{static} network, each component represents an independent group of applicants. The links in a component represent the collaborations between its applicants. Through these links, diffusion and flow of knowledge can be realized and, consequently, applicants can influence each other, either directly or indirectly. Nonetheless, knowledge is trapped within the bounds of a component, therefore, the more fragmented a network is, the less it can facilitate collaboration and knowledge spreading. In a \textit{growing} network, the addition of the critical patent at the percolation threshold  invokes the birth of the Giant Component, which marks the beginning of the transformation of the system from a sea of small-sized, independent components - islands of “localized” patent-induced collaborations - to a tangible network that forms the grounds for more “globalized” collaborations. Thus, it is of great interest to \textbf{\textit{study the circumstances that pertain to the GC formation and how these change in time}}.

Performing a temporal analysis on the GC-formation specifics would enable statistical inference and the detection of patterns or trends likely to reveal useful information about its history and evolution. To this end, we employed the “time-window of sliding origin” concept, which allows us to exploit all of the available 35 years of patents data. This concept enables us to study aspects of the GC formation like the variation of time and number of patents required for reaching the percolation threshold, as well as certain characteristics of the two major groups of applicants, immediately before their union, i.e. immediately before the birth of the GC.

\subsection{The "time-window of sliding origin" concept and the GC formation throughout the timeline}

By starting the temporal reconstruction of the system from June 1978\footnote{The earliest date in the data.} we make the tacit - and erroneous - assumption that no patents were filed before this date. Nonetheless, this false assumption allows us to open up a “window” on the temporal evolution of the system and examine the circumstances pertaining to the emergence of the GC \textbf{\textit{at this particular point in the data timeline}}.

Clearly, one could start the reconstruction from any arbitrarily chosen date and open up a new window from that date to study the GC formation. Thus, by progressively moving forward (sliding) the starting date (origin) of the time-window, we end up with a set of time-windows that form a chain of snapshots of the evolving system. This notion embodies the “time-window of sliding origin” concept that permits us to exploit all of the available 35 years of data and perform a temporal analysis on the circumstances under which the GC forms, \textbf{\textit{throughout the available time span}}.

\begin{figure}[ht]
    \centering
    \includegraphics[width=0.6\textwidth]{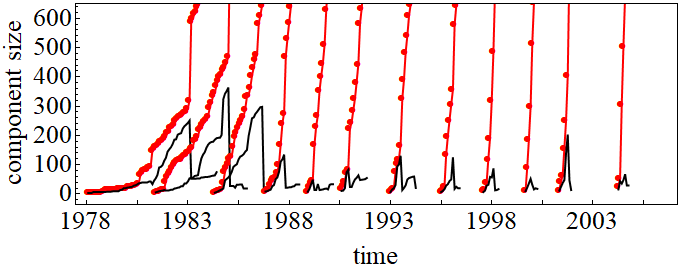}
    \caption[Damage report!]{The emergence of the GC, in different points in time. Each GC corresponds to a different “time-window”, opened up on the system’s timeline. (Red circles: Largest Connected Component (LCC), black lines: Second Largest Connected Component (SLCC)). \label{fig:LCgrowthDetailWindows}}
\end{figure}

Fig.~\ref{fig:LCgrowthDetailWindows} depicts the emergence of the GC captured on a sample group of time-windows\footnote{All percolation thresholds were confirmed by the method described in \ref{sec:percConfirm}.}. It is evident that the time required for the GC formation varies with the starting date of the window. Specifically, it appears that the GC takes significantly longer to form in the early windows than in the windows that start later in the timeline. 

\subsection{Variation of time and patents required for the GC formation through time}

\begin{figure}[ht]
    \centering
    \includegraphics[width=0.9\textwidth]{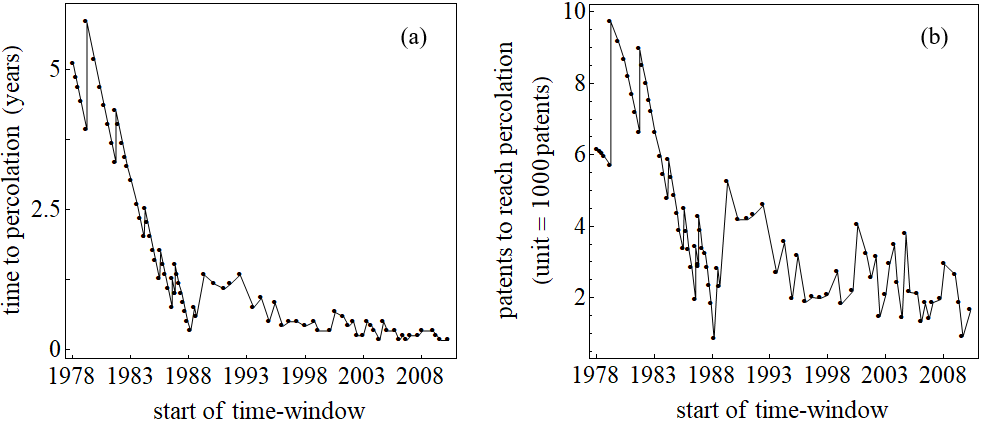}
    \caption[Damage report!]{(a) Elapsed time to percolation and (b) patents filed until percolation vs. starting date of sliding time-window. \label{fig:maziYearsIndices}}
\end{figure}

The findings of Fig.~\ref{fig:LCgrowthDetailWindows} clearly warrants a more detailed analysis. Therefore, we examined the time required to reach the percolation threshold on a larger set of time-windows densely distributed in the timeline of the data. The results are shown in Fig.~\ref{fig:maziYearsIndices}(a), which depicts the \textit{time elapsed between the starting date of each time-window and the date on which the GC forms vs. the starting date of the time-window}.

It is evident that it takes a considerably larger amount of time for the GC to form when the origin of the time-window falls into the early years of the timeline, i.e. roughly the first decade (late 70s to late 80s). Regarding the remaining years (late 80s to early 10s), there is an indication of further division into two more regimes (late 80s to middle 90s and middle 90s to early 10s), albeit a weaker one. The results for the number of patents required to reach the percolation threshold are very similar, as shown in Fig.~\ref{fig:maziYearsIndices}(b).

\subsection{Qualitative differences during the growth process}

The system’s behaviour, during the growth phase leading up to percolation, also reveals qualitative variations that separate the timeline into at least two - possibly three - regimes. This phase was portrayed in detail in Fig.~\ref{subsec:GCemergence}, for the time-window whose origin is the first date of the data. In that case, the percolation transition requires a considerable amount of time (circa 5 years), during which the two largest evolving components grow independently of each other, until the critical patent enters the system and they merge abruptly, Fig.~\ref{fig:LCgrowthDetail}. The resulting component is the network’s GC, which continues to grow and quickly becomes vastly larger than any other component. As mentioned before, similar descriptions of this growth process have been previously recorded~\cite{newman2001structure,bettencourt2009scientific,liu2015evolutionary,liu2015structure}. The exact same qualitative behaviour is observed in all time-windows with starting dates within the early period (late-70s to late-80s, Fig.~\ref{fig:LCgrowthDetailWindows}).

However, this regularity breaks down during the second period (late-80s to mid-90s), as time-windows that reveal qualitatively different behaviour are noticed sporadically for the first time. In those windows, it appears that the LCC merges with the SLCC\footnote{Each time the LCC merges with the SLCC, the component which at that time ranks third (with respect to its size) becomes the new SLCC of the network.} on more than one occasion, long before the merge that brings on the actual percolation, when both components are still small-sized. As a consequence, when the percolation threshold is reached, the LCC is already noticeably larger than any other component and therefore the percolation transition is rather unremarkable.

Overall, the evolving system’s behaviour regarding the percolation transitions, in the three time-periods, can be described as follows: During the first period all percolation transitions are “clean” and unambiguous and require considerable amounts of time. In the second and third period, the percolation transitions are typically shorter than those of the first period. Additionally, the second period introduces time-windows, which feature step-wise and therefore less dramatic percolation transitions. These gradual transitions become even more frequent during the third period. Given that we were mainly interested in comparing the characteristics of the two groups of applicants represented by the two largest components at the percolation threshold, we only considered time-windows with marked percolation transitions at which the two largest components are of similar size, in all subsequent calculations. At any rate, these windows are densely distributed throughout the timeline and therefore there is no loss of generality.

\subsection{Characteristics of the two major groups of applicants through time: technological correspondence and geographical overlap}

Three features of the GC formation process have already been examined and were found to change with time. These are: the amount of time and patents required for the GC to emerge (Fig.~\ref{fig:maziYearsIndices}) and – from a qualitative standpoint – the percolation transitions and the growth process leading up to it.

Further information can be extracted by investigating and comparing the characteristics of the two major groups of applicants that correspond to the LCC and SLCC of the growing network and are ultimately united by the critical (red-bond) patent into the GC, at the percolation threshold. We define the “adjacent-pre-percolation state” as the state of the LCC and SLCC immediately before percolation and ask the following question: \textbf{\textit{can we identify any features that facilitate/inhibit the union of the two major groups of applicants, i.e. the GC formation, by studying the characteristics of the LCC and SLCC in the “adjacent-pre-percolation state”, over time?}} To answer that question we compared the two groups of applicants (LCC and SLCC) in their “adjacent-pre-percolation state”, in terms of a. the technological areas of the patents involved and b. the geographical origin of the applicants, in multiple time-windows.

\subsubsection{Technological proximity of the two major GC groups of applicants (LCC and SLCC in the adjacent-pre-percolation state) over time}

To assess the technological proximity of the two groups, we utilized the International Patent Classification (IPC) codes~\cite{wipoIPC} of the patents. The IPC codes are used by the European Patent Office (EPO) as a means for classifying the patents with respect to their relative technological areas and conversely, have been used as a tool for determining the technological areas of a patent~\cite{choi2015predictive,jun2014small,park2015network}. There are four types of IPC categorization (from the most coarse-grained to the most refined): Section, Class, Subclass and Group (Main and Subgroup). Each of these categorizations features a set of IPC codes. The Section IPC categorization was used in all calculations, as it is the most coarse-grained of all, with many patents falling into each of the eight sections and therefore yields the most (statistically) reliable results. The databases used in this study provide the 8th edition IPC codes, for each patent. 

We determined the frequency distribution of the eight IPC sections for the two groups of applicants (LCC and SLCC in the adjacent-pre-percolation state) that join to form the GC, for multiple time-windows. The degree of similarity/dissimilarity of these distributions serves as a measure of the technological proximity between these two groups. It was found that the technological proximity varies over time (Fig.~\ref{fig:sectionsInTime} in ~\ref{sec:SupplFigs}). Specifically, in most of the early period time-windows the distributions overlap significantly, in the middle period they tend to be less comparable and finally in the majority of the late period windows they tend to differ considerably.

In order to quantify the technological proximity of the two groups of applicants, we employed the normalized Euclidean distance, $d$, between the two 8-dimensional vectors u and v corresponding to the IPC sections frequencies of the two groups: 

\begin{equation}
\label{tech_distance}
d = \frac{1}{2}\frac{|u^{'}-v^{'}|^{2}}{|u^{'}|^{2}+|v^{'}|^{2}} 
\end{equation}

where $u^{'}=u-\bar{u}$, $v^{'}=v-\bar{v}$, $\bar{u}=\frac{1}{N}\sum_{i=1}^{N} u_i$, $\bar{v}=\frac{1}{N}\sum_{i=1}^{N} v_i$ and $N = 8$, the eight IPC sections. 

\begin{figure}
    \centering
    \begin{minipage}{0.49\textwidth}
        \centering
        \includegraphics[width=0.9\textwidth]{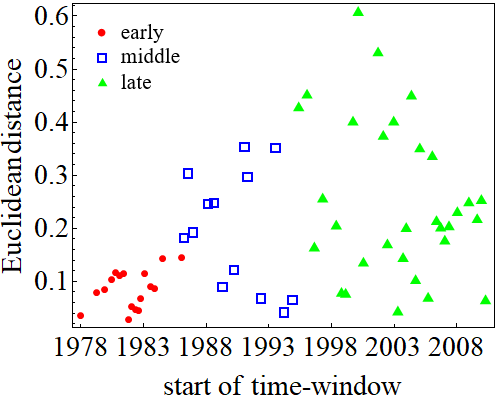} 
        \caption[Damage report!]{Normalized Euclidean distance of the IPC codes distribution, of the two major components that make up the GC, in their adjacent-pre-percolation state for multiple time-windows in the available timeline. Red disks, blue squares and green triangles are used to visualize a rough division of the timeline into the early (1978-1986), middle (1986-1995) and late (1995-2013) periods of the timeline.\label{fig:euclideanInTime}}
    \end{minipage}\hfill
    \begin{minipage}{0.49\textwidth}
        \centering
        \includegraphics[width=0.9\textwidth]{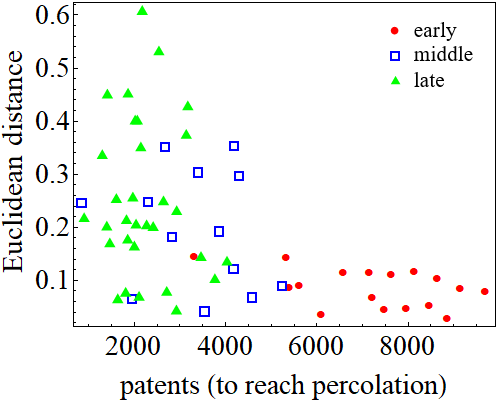} 
        \caption[Damage report!]{Normalized Euclidean distance of the IPC codes distribution, of the two major clusters that make up the GC, in their adjacent-pre-percolation state, vs. the required number of patents to reach the percolation state, for multiple time-windows in the available timeline. Red, blue and green colour is used to visualize a rough division of the timeline into the early (1978-1986), middle (1986-1995) and late (1995-2013) periods. \label{fig:euclideanVsPatents}}
    \end{minipage}
\end{figure}

The calculated Euclidean distance, depicted in Fig.~\ref{fig:euclideanInTime}, varies over time in a way that is once more suggestive of a three-period division of the timeline. This division is by no means clear-cut, nonetheless, it is evident that high technological distance between the two components became progressively more probable through the three periods. A plausible approximate partition would be 1978-1986, 1986-1995 and 1995-2013, shown in Fig.~\ref{fig:euclideanInTime} in red, blue and green, respectively, for visualization purposes. This provides us with a rough estimation of the Euclidean distance in the three periods: from an average of 0.086 with a very low variance (0.001) in the early windows, to 0.196 with a higher than tenfold leap in its variance (0.013) during the middle ones, to 0.253, which is almost three times higher than that of the early period and with a higher still variance (0.022), in the late windows.

Both the number of patents required to reach percolation (Fig.~\ref{fig:maziYearsIndices}b) and the technological distance of the two major components in the adjacent-pre-percolation state (Fig.~\ref{fig:euclideanInTime}) have been shown to change over time in a manner that separates the timeline into three regimes. To probe into the relation between these two quantities, the two figures were combined into one, Fig.~\ref{fig:euclideanVsPatents}, by eliminating the common variable, i.e. the time. We used the same approximate partition of the timeline as in Fig.~\ref{fig:euclideanInTime} to denote the period of each time-window (early, middle and late).

In Fig.~\ref{fig:euclideanVsPatents}, the points representing time-windows of the same region tend to cluster together, which further supports the findings of Fig.~\ref{fig:maziYearsIndices} and ~\ref{fig:euclideanInTime} regarding the division of the timeline into at least two - if not three - regimes. Moreover, Fig.~\ref{fig:euclideanVsPatents} hints at an association between the two factors (the technological proximity of the two groups of applicants and the amount of patents required for the GC formation), which is strong for the time-windows of the early period and quite weaker for the rest of the timeline. In the early period time-windows, requiring a high amount of patents to reach percolation most likely co-exists with low Euclidean distance (high technological proximity) between the two major groups of applicants that constitute the GC. For the rest of the timeline, the lower the amount of patents the more likely high technological distance becomes, however there is still a good chance for low technological proximity in many time-windows of low amount of patents. Overall, as the system evolves, it appears that it undergoes notable changes that are reflected in both these factors pertaining to the GC formation. Namely, the \textbf{\textit{average}} value of both quantities shifts, as the system moves through the three regimes, according to the following pattern: from a period characteristic of low Euclidean distance and high number of patents, to a period of medium-high Euclidean distance and number of patents, and, finally, to a period of higher Euclidean distance and lower number of patents.

Similar results were obtained with the use of Pearson correlation coefficient, r, as a measure of the technological proximity between the two components that form the GC, when plotted against the number of patents required to reach percolation, Fig.~\ref{fig:pearsonVsPatents} in ~\nameref{sec:SupplFigs}. The null hypothesis is that the two sets (of IPC codes distribution) are independent, with the significance level set to 0.05.

\subsubsection{Geographical interplay between the two major groups of applicants (LCC and SLCC in the adjacent-pre-percolation state) over time}

The geographical breakdown of the applicants in both groups in the adjacent-pre-percolation state, for multiple time-windows, revealed a series of interesting facts. First, the bulk of the applicants in both groups were found to be operating in just three countries, namely, France (FR), Germany (DE) and Japan (JP), at all times. Applicants from the United States (US) are also present, with a much lower yet non-negligible contribution. Second, for all time-windows investigated, one of the two groups mostly comprises FR\/DE applicants, while the other of JP ones. Third, the level of geographical confinement of the two groups appears to change over time, in a way that is once again implying a three-region division of the timeline (Fig.~\ref{fig:applicantCountryCountEML} in ~\nameref{sec:SupplFigs}). 

\begin{figure}[ht]
    \centering
    \includegraphics[width=0.8\textwidth]{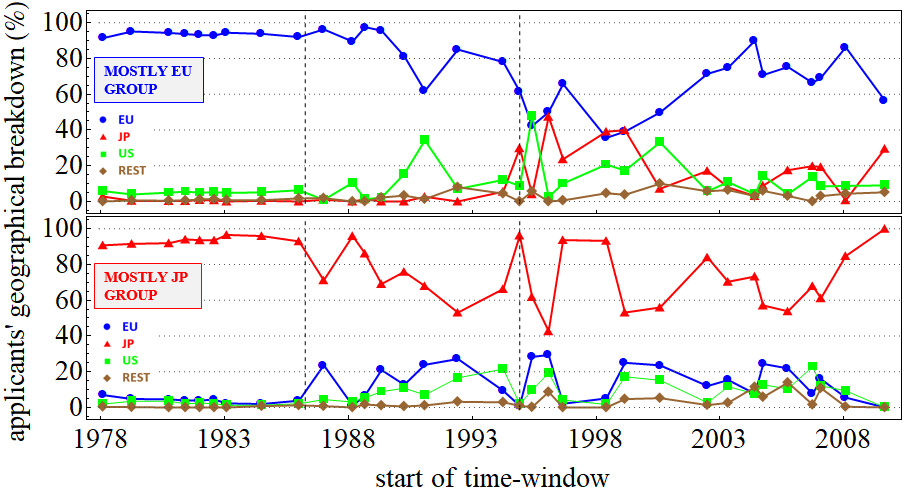}
    \caption[Damage report!]{Geographical breakdown of the groups of applicants corresponding to the LC and SLC, as captured exactly before the addition of the critical bond which joins them into the GC (adjacent-pre-percolation state). At (almost) all times, one group consists of mostly EU applicants, while the other of mostly JP. The depicted rough division of the timeline (two vertical lines) is the same as in Fig.~\ref{fig:euclideanInTime} and ~\ref{fig:euclideanVsPatents} (EU: blue disks, JP: red triangle, US: green squares, rest of the world: brown rhombi) \label{fig:groupsGeographyInTime}}
\end{figure}

In light of these findings, we classified the country occurrences into four categories: all European (EU), Japan (JP), United States (US) and all the rest countries in the world (REST). The percentages of these four categories, calculated for both groups, illustrate the geographical interplay between the two applicants’ groups. The results, shown in Fig.~\ref{fig:groupsGeographyInTime}, corroborate the three-part aforementioned division of the 35-year timeline.

During the first regime (ca. 1978-1986), the applicants in the two groups are clearly segregated, confined either in EU or JP, while US and rest of the world applicants are practically absent. Unlike the first regime that is effortlessly distinguishable from the rest, the distinction between the second (ca. 1986-1995) and the third (ca. 1995-2013) regimes is significantly less prominent. Both regimes exhibit geographical interactions between the two groups, in stark contrast to the complete segregation witnessed during the first regime.  The intensity of these interactions however is much lower in the second regime. In particular, while there is a notable rise in both EU and US percentages in the “mostly JP” group, only the US percentage rises in the mostly “EU” group; Japanese applicants are yet absent from this group during the second period. Lastly, applicants from the rest of the world join in the network for the first time in the last two regimes. The contribution of these applicants is larger in the third regime in both groups, nonetheless it remains rather small at all times. 

\begin{figure}[ht]
    \centering
    \includegraphics[width=0.7\textwidth]{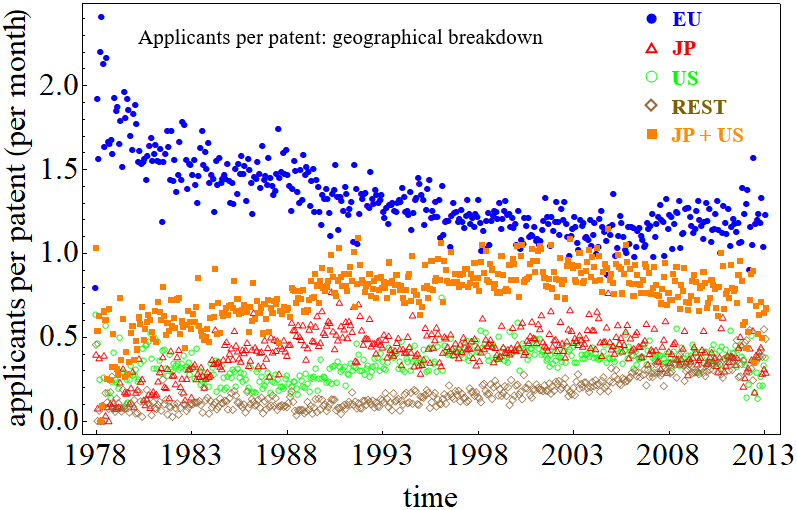}
    \caption[Damage report!]{Geographical breakdown of the number of the applicants over submitted patents, per month, throughout the timeline. (EU: blue disks, JP: red triangles, US: green circles, rest of the world: brown rhombi, JP+US: orange squares) \label{fig:applicantGeogrBreakdown}}
\end{figure}

These results prompted a supplementary statistical analysis on the geography of the raw patent data from which the network examined so far was derived. As mentioned before, this is the data set that comprises all patents, excluding the ones with just one applicant\footnote{These patents result in isolated (island) nodes or self-loops. For results for the network with one-applicant patents included see Fig.~\ref{fig:averageApplicantPerPatentDS_A} in ~\nameref{sec:SupplFigs}.}. First, we determined the monthly average of the number of applicants\footnote{Given that these are the raw patent data, the applicants in this analysis are not unique, since an applicant can participate in multiple patents, throughout the timeline.} per patent (Fig.~\ref{fig:averageApplicantPerPatentDS_B} in ~\nameref{sec:SupplFigs}), which remains remarkably constant, averaging ~2.25 throughout the timeline. This quantity was subsequently broken down into the four aforementioned geographical categories, EU, JP, US and the rest of the world (REST), in Fig.~\ref{fig:applicantGeogrBreakdown}.

During the first ~12-14 years of the timeline, there is a sharp decline in the EU applicants, which coincides with a corresponding rise in the JP applicants. During the same period, the number of US applicants is pivoting around a low but non trivial value, while the REST applicants barely exist. For the next ~15 years, US and JP applicants participate at approximately the same - quite stable - rate, while the EU applicants continue to decline, albeit at a much slower rate, which now coincides with a slow but steady rise in the REST applicants. Finally, for the last ~5-6 years the US and JP applicants count fall just enough for the rest of the world applicants to reach them, whose participation continues its steady rise. Overall, it appears that non-European applicants’ participation in the EPO patents increases over time, which leads to more opportunities for the formation of inter-continental collaborations.

\begin{figure}[ht]
    \centering
    \includegraphics[width=0.6\textwidth]{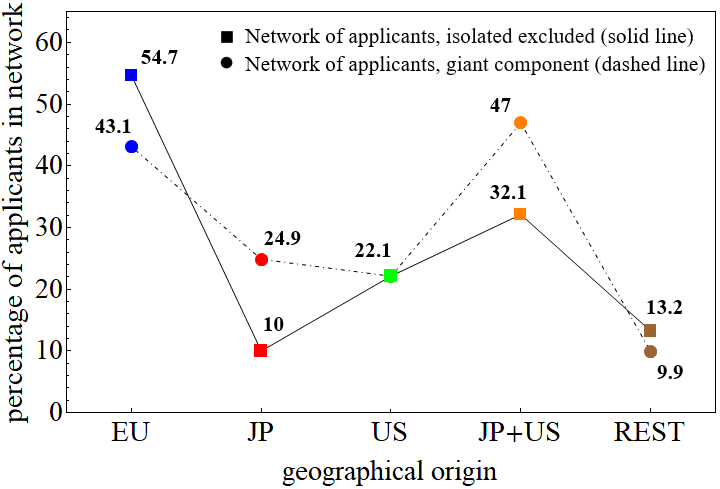}
    \caption[Damage report!]{Geographical breakdown of the applicants in the network resulting from the time-aggregated data, spanning all available timeline. (Squares: the network of applicants, isolated nodes excluded, Disks: the GC of the network) \label{fig:applicantsInNetGeog}}
\end{figure}

Finally, to complete this geographical analysis, we look into the static network, i.e. the network that results from all the patents aggregated over the whole 35 years of data, excluding only the patents with just one applicant\footnote{As mentioned before, these patents result in isolated (island) nodes or self-loops. For results for the network with these patents included see Fig~\ref{fig:applicantsInNetGeogWL2} in ~\nameref{sec:SupplFigs}.}. The geographical breakdown of the applicants in that network and its GC is shown in Fig.~\ref{fig:applicantsInNetGeog}. 

Fig.~\ref{fig:applicantsInNetGeog} complements the findings of Fig.~\ref{fig:groupsGeographyInTime} and ~\ref{fig:applicantGeogrBreakdown} regarding JP as it corroborates the crucial part that it plays in the core of the EPO patents network. Unlike the US applicants that submit patent applications indiscriminately, JP applicants principally participate in patent collaborations that contribute to the GC of the network.

\subsection{Are the major patent-contributors more likely to participate in a critical patent?}\label{sec:applicantsRank}

Last but not least, we explored the question of whether the likelihood of participating in a critical patent increases with the total amount of patents that an applicant contributes to the network or with its size. The latter is also related to the firm size vs. innovation output problem~\cite{schumpeter1934theory,schumpeter1942socialism} (see~\nameref{sec:intro}).

To this end, we ranked the applicants with respect to their overall patents production, over several time-windows distributed in the 35-year timeline and assessed the relative position of applicants participating in the critical-patents. Thus, for each time-window, we brought the network into the percolation threshold state and determined the percentage of the patents for all applicants having filed at least one patent\footnote{All patents are included in this analysis, even those with only one applicant.} from the starting date of the time-window to the percolation threshold. For each time-window we calculated the percentage $p_a$ for applicant $a$ according to the formula:

\begin{equation}
\label{pat_percent}
p_a = \frac{patents \ filed \ by \ applicant \ a \ in \ t_p}{total \ of \ patents \ filed \ in \ t_p} \ x \ 100
\end{equation}

where $t_p$ is the time elapsed from the starting date of the time-window to the date of the critical patent (percolation threshold). We subsequently evaluated the applicants' rank and percentile rank, according to their patent percentage, which are shown in Table~\ref{tab:rank} for the red-bond firms of nine representative time-windows in the data timeline. 

The firms in the top 1\% were found to submit nearly half the patents ($\sim$ 48\%) on average, yet, they participate in the red-bond patents on an average of only $\sim$ 24\%. Therefore, we conclude that the top patent-producing applicants do not dominate the production of red-bond links that bring the GC into existence and induce the network’s coherence, as one would expect, with respect to their overall patent output. Our results suggest that the reason behind this discrepancy is that many of the firms in this category (the top 1\%), have a low percentage of collaborative patents. Thus, the likelihood of them yielding a critical patent is reduced in spite of them being highly productive on the whole. 

Finally, the plausible assumption that the larger a firm's size the more patents it is capable of producing~\cite{tether2002co} is overall corroborated by our results, as the bulk of the top patent-producing applicants are large-sized ones. Additionally, the majority of the red-bond applicants (Bosch, Toyota, Hitachi, Honda, Nissan, Phillips, to name a few) in this network are found to be large-sized firms, regardless of their total patent output in each time-window. 

These findings have further implications re the firm size vs. innovation-output debate (see~\nameref{sec:discussion}).

\begin{table}[ht]
 \caption{ \label{tab:rank}
  Rank and percentile rank according to patent percentage, for the red-bond firms of nine representative time-windows in the data timeline.}
 \centering
 \begin{tabular}{c c c c c }
  \hline \hline
  year of time-window & critical patent applicants & rank & percentile (\%) \\
  \hline
  \multirow{2}{*}{1978} & \textit{applicant 1} & 17 & 99.75 \\
  \textsc{} & \textit{applicant 2} & 115 & 98.30 \\
   \hline
  \multirow{5}{*}{1983} & \textit{applicant 1} & 77 & 98.91 \\
  \textsc{} & \textit{applicant 2} & 87 & 98.77 \\
  \textsc{} & \textit{applicant 3} & 91 & 98.71 \\
  \textsc{} & \textit{applicant 4} & 374 & 94.70 \\
  \textsc{} & \textit{applicant 5} & 2044 & 71.02 \\
  \hline
  \multirow{2}{*}{1989} & \textit{applicant 1} & 2 & 99.93 \\
  \textsc{} & \textit{applicant 2} & 10 & 99.64 \\
  \hline
  \multirow{2}{*}{1993} & \textit{applicant 1} & 38 & 90.28 \\
  \textsc{} & \textit{applicant 2} & 521 & 99.64 \\
  \hline
  \multirow{2}{*}{1997} & \textit{applicant 1} & 29 & 98.99 \\
  \textsc{} & \textit{applicant 2} & 586 & 79.50 \\
  \hline
  \multirow{2}{*}{2000} & \textit{applicant 1} & 189 & 96.60 \\
  \textsc{} & \textit{applicant 2} & 584 & 89.49 \\
  \hline
  \multirow{2}{*}{2004} & \textit{applicant 1} & 58 & 97.09 \\
  \textsc{} & \textit{applicant 2} & 109 & 94.54 \\
  \hline
  \multirow{2}{*}{2007} & \textit{applicant 1} & 5 & 99.80 \\
  \textsc{} & \textit{applicant 2} & 249 & 90.01 \\
  \hline
  \multirow{2}{*}{2010} & \textit{applicant 1} & 71 & 94.95 \\
  \textsc{} & \textit{applicant 2} & 71 & 94.95 \\
  \hline \hline
 \end{tabular}
\end{table}

\section{Discussion}\label{sec:discussion}

Overall, our results reveal that as the patent network evolves through time, the percolation threshold (on average) comes at shorter times, requires fewer patents, features increasing inter-regional collaborations and increasing technological distance between the two major groups of applicants involved. How do all these fit together?

It is plausible to presume that the basic underlying factor in this change is the rise in the inter-continental collaborations and the resulting globalization, through the decades. Thus, in the early years -  when advanced communication technologies were not yet widespread - geographical distance was most likely hindering inter-continental collaborations and therefore the two groups were segregated for many years until their eventual union. It is likely that this is the reason behind the high technological similarity between the two groups as they were forced to function independently for long periods of time. During the transitional phase of the middle years, a change develops that appears to mitigate the negative effect of distance on inter-continental collaborations. Therefore, windows of shorter time and higher technological distance between the two groups start to emerge. Lastly, the shortest intervals to percolation and higher technological distance are seen recurrently in late-years windows, when geographical interplay is at its highest. 

This conclusion is in agreement with similar results found by studying patent collaborations with statistical methods~\cite{ma2008patent,guan2012patent}. A number of factors could be identified as plausible contributors to this globalization phenomenon, such as the vastly available air-travel, the remarkable upsurge in the use of internet and various other socio-economic factors that forced the applicants to reach out further to build complementing collaborative ties and meet the demands of an increasingly competitive and fast-paced world.

Moreover, regarding the firm size vs. innovation-output debate, our results imply that large firm size favours innovation activities. First, the assumption that the top patent-producing applicants are also likely to be large-sized ones~\cite{tether2002co} was corroborated by our analysis. Second, the majority of the red-bond applicants (Bosch, Toyota, Hitachi, Honda, Nissan, Phillips, to name a few) in this network were found to be large-sized firms regardless of the total amount of patents they contribute in each time-window. Since a patent is in itself an innovation output indicator and furthermore, the critical bonds of a network are believed to correspond to a highly innovative events~\cite{bettencourt2009scientific}, we infer that in the EPO patents network, it is the large-sized applicants (irrespective of the magnitude of their total patent contribution) that are most likely to introduce radical innovation. Therefore our findings are suggestive of a positive correlation between size and innovation output, as in the majority of the studies reviewed in~\cite{becheikh2006lessons}, as well as in~\cite{schumpeter1942socialism,tether1998small}.

Lastly, the striking presence of Japan in the EPO patent network should not go unnoticed. In all time-windows examined, the Japanese applicants make up a significant portion - nearly half - of the GC, at the moment of its emergence, i.e. at the percolation threshold (Fig.~\ref{fig:groupsGeographyInTime}). Furthermore, the fact that Japan seems to forge strategic alliances that are key to the network's coherence also surfaced in two more of our findings. Firstly, Japan's participation in the static, time-aggregated network never exceeds 10\% (see Fig.~\ref{fig:applicantsInNetGeog} and Fig.~\ref{fig:applicantsInNetGeogWL2}), while it reaches 25\% in its GC. Secondly, an abundance of Japanese applicants (Toyota, Hitachi and subsidiaries, Honda, Nissan, Ube industries, JSR corporation, Riken, the National Institute of Advanced Industrial Science And Technology etc.) consistently appear in red-bond patents, namely the patents that induce the GC emergence. Interestingly enough, judging from the type of activities of these companies and from the red-bond patents themselves, it appears that they are all related, in one way or another, to the automobile industry. Thus, we believe that the automobile industry has played a crucial part in the EPO patent network and consequently to the introduction and - even more so - to the diffusion of innovation.

\section{Conclusions}

We performed a temporal analysis, on the patent applicants’ collaboration network derived from the REGPAT patents data, which span 35 years, from year 1978 (June) to 2013 (July). Specifically, we studied the network over a collection of time-windows, which allowed us to exploit the whole timeline of 35 years and study the evolution of the applicant’s collaboration network dynamically, instead of limiting the investigation to the aggregated, static network. This approach enabled us to open up multiple windows of observation onto the system’s evolution, with starting dates distributed throughout the timeline. We focused our analysis on the GC, and specifically on certain characteristics of its formation, i.e. the network’s percolation, in multiple time-windows. This analysis, uncovered evidence of qualitative and quantitative differentiations of characteristics such as, the amount of time/patents required for percolation, the technological similarity and the geographical overlap of the major groups of applicants which make up the GC and the abruptness of the percolation transition. All these aspects of the GC formation are found to change in time in a way that is suggestive of a three-regime division of the timeline. 

Specifically, during the first period (ca. 1978-1987, early-years), we observe clear-cut percolation transitions, in which the two largest, separately growing components are merged abruptly into the network’s GC. The groups of applicants represented by these two components are found to exhibit high technological resemblance, based on the IPC codes of their corresponding patents. Furthermore, in all early-years windows, the two groups are almost completely geographically segregated into two regions, EU (principally FR and DE) and JP, with a very small, but nonetheless measurable US fraction of applicants being present in both communities. Moreover, this period is marked by the highest amounts of time/patents required for the GC formation. 

During the second period (ca. 1987-1995, middle years), there is a shift in the percolation transition process, in many windows. In such windows, the transition appears to be shorter in time, more gradual and less striking. Overall, during this period, the two groups of applicants that make up the GC are more complementary and less similar technologically, and begin to exhibit some geographical overlap. Specifically, the community comprising mostly JP applicants, starts to welcome collaboration with both the EU and the US, however, the respective EU community still remains mostly detached. Additionally, the amounts of both time and patents required for the GC formation lessens, on average. 

In the third period (ca. 1995-2013, late years), even more windows display the less-abrupt, less outstanding percolation transition behaviour. Overall, it is the period in which the two groups of applicants demonstrate the greatest technological distance, as well as the greatest geographical overlap. There is a notable rise in both frequency and intensity of collaborations between Europe, Japan and the US, in both groups. Furthermore, in this period the windows displaying the least amount of time/patents required for percolation have become recurrent.

Moreover, our results indicate that top patent-producing applicants are likely to yield a critical patent, however at a rate significantly lower than their overall patent production. Additionally, the top patent-producing applicants are predominantly large-sized firms, supporting the assumption that the largest the firm size the more patents it is capable of producing\cite{tether2002co}. Finally, the red-bond applicants (Bosch, Toyota, Hitachi, Honda, Nissan, Phillips amongst others) are also large-sized firms, notwithstanding the amount of patents they contribute in each time-window. Therefore, regarding the GC formation and the network’s coherence, the significant applicants are typically large-sized ones, but not necessarily amongst the top patent-producing.

Last but certainly not least, essentially all of our findings stress the vital importance of Japan to the GC of the EPO patent network. Firstly, in all time-windows, one of the two major groups that join into the GC consists mainly of Japanese applicants. Secondly, Japanese applicants make up one quarter of the static, aggregated network’s CG (while they constitute just 10\% of the whole network). Thirdly, Japanese firms and research institutes are found to systematically participate in red-bond patents. Toyota, Hitachi and subsidiaries, Honda, Nissan, Ube industries, JSR corporation, Riken, the National Institute of Advanced Industrial Science And Technology among others are consistently featured in the critical patents.

\section*{Acknowledgements}

This work was supported
by the European Commission FET Project MULTIPLEX
No. 317532. We thank Michael Kanetidis for his valuable comments.

\bibliographystyle{unsrt}  
\bibliography{references} 

\begin{thebibliography}{10}

\bibitem{solow1957technical}
Robert~M Solow.
\newblock Technical change and the aggregate production function.
\newblock {\em The review of Economics and Statistics}, pages 312--320, 1957.

\bibitem{griliches1958research}
Zvi Griliches.
\newblock Research costs and social returns: Hybrid corn and related
  innovations.
\newblock {\em Journal of political economy}, 66(5):419--431, 1958.

\bibitem{mansfield1961technical}
Edwin Mansfield.
\newblock Technical change and the rate of imitation.
\newblock {\em Econometrica: Journal of the Econometric Society}, pages
  741--766, 1961.

\bibitem{mansfield1962entry}
Edwin Mansfield.
\newblock Entry, gibrat's law, innovation, and the growth of firms.
\newblock {\em The American economic review}, 52(5):1023--1051, 1962.

\bibitem{griliches1964research}
Zvi Griliches.
\newblock Research expenditures, education, and the aggregate agricultural
  production function.
\newblock {\em The American Economic Review}, 54(6):961--974, 1964.

\bibitem{mansfield1965rates}
Edwin Mansfield.
\newblock Rates of return from industrial research and development.
\newblock {\em The American Economic Review}, 55(1/2):310--322, 1965.

\bibitem{griliches1980returns}
Zvi Griliches.
\newblock Returns to research and development expenditures in the private
  sector.
\newblock In {\em New developments in productivity measurement}, pages
  419--462. University of Chicago press, 1980.

\bibitem{griliches1983comparing}
Zvi Griliches and Jacques Mairesse.
\newblock Comparing productivity growth: an exploration of french and us
  industrial and firm data.
\newblock {\em European Economic Review}, 21(1-2):89--119, 1983.

\bibitem{griliches1985productivity}
Zvi Griliches.
\newblock Productivity, r\&d, and basic research at the firm level in the
  1970s.
\newblock Technical report, National Bureau of Economic Research, 1985.

\bibitem{NAP612}
Ralph Landau and Nathan Rosenberg, editors.
\newblock {\em The Positive Sum Strategy: Harnessing Technology for Economic
  Growth}.
\newblock The National Academies Press, Washington, DC, 1986.

\bibitem{romer1986increasing}
Paul~M Romer.
\newblock Increasing returns and long-run growth.
\newblock {\em Journal of political economy}, 94(5):1002--1037, 1986.

\bibitem{geroski1989entry}
Paul~A Geroski.
\newblock Entry, innovation and productivity growth.
\newblock {\em The Review of Economics and Statistics}, pages 572--578, 1989.

\bibitem{caballero1993high}
Ricardo~J Caballero and Adam~B Jaffe.
\newblock How high are the giants' shoulders: An empirical assessment of
  knowledge spillovers and creative destruction in a model of economic growth.
\newblock {\em NBER macroeconomics annual}, 8:15--74, 1993.

\bibitem{geroski1993profitability}
Paul Geroski, Steve Machin, and John Van~Reenen.
\newblock The profitability of innovating firms.
\newblock {\em The RAND Journal of Economics}, pages 198--211, 1993.

\bibitem{klette1996r}
Tor~Jakob Klette.
\newblock R\&d, scope economies, and plant performance.
\newblock {\em The RAND Journal of Economics}, pages 502--522, 1996.

\bibitem{crepon1998research}
Bruno Cr{\'e}pon, Emmanuel Duguet, and Jacques Mairessec.
\newblock Research, innovation and productivi [ty: An econometric analysis at
  the firm level.
\newblock {\em Economics of Innovation and new Technology}, 7(2):115--158,
  1998.

\bibitem{harhoff1998r}
Dietmar Harhoff.
\newblock R\&d and productivity in german manufacturing firms.
\newblock {\em Economics of Innovation and New Technology}, 6(1):29--50, 1998.

\bibitem{klette2000accumulation}
Tor~Jakob Klette and Frode Johansen.
\newblock Accumulation of r\&d capital and dynamic firm performance: a
  not-so-fixed effect model.
\newblock In {\em The Economics and Econometrics of Innovation}, pages
  367--397. Springer, 2000.

\bibitem{klomp1999importance}
Luuk Klomp and George Van~Leeuwen.
\newblock The importance of innovation for company performance.
\newblock {\em Netherlands Official Statistics}, 14(2):26--35, 1999.

\bibitem{loof2002knowledge}
Hans L{\"o}{\"o}f and Almas Heshmati.
\newblock Knowledge capital and performance heterogeneity:: A firm-level
  innovation study.
\newblock {\em International Journal of Production Economics}, 76(1):61--85,
  2002.

\bibitem{cainelli2004impact}
Giulio Cainelli, Rinaldo Evangelista, and Maria Savona.
\newblock The impact of innovation on economic performance in services.
\newblock {\em The Service Industries Journal}, 24(1):116--130, 2004.

\bibitem{parisi2006productivity}
Maria~Laura Parisi, Fabio Schiantarelli, and Alessandro Sembenelli.
\newblock Productivity, innovation and r\&d: Micro evidence for italy.
\newblock {\em European Economic Review}, 50(8):2037--2061, 2006.

\bibitem{hall2009innovation}
Bronwyn~H Hall, Francesca Lotti, and Jacques Mairesse.
\newblock Innovation and productivity in smes: empirical evidence for italy.
\newblock {\em Small business economics}, 33(1):13--33, 2009.

\bibitem{cassiman2010innovation}
Bruno Cassiman, Elena Golovko, and Ester Mart{\'\i}nez-Ros.
\newblock Innovation, exports and productivity.
\newblock {\em International Journal of Industrial Organization},
  28(4):372--376, 2010.

\bibitem{hall2011innovation}
Bronwyn~H Hall.
\newblock Innovation and productivity.
\newblock Technical report, National bureau of economic research, 2011.

\bibitem{mohnen2013innovation}
Pierre Mohnen and Bronwyn~H Hall.
\newblock Innovation and productivity: An update.
\newblock {\em Eurasian Business Review}, 3(1):47--65, 2013.

\bibitem{medda2014technological}
Giuseppe Medda and Claudio~A Piga.
\newblock Technological spillovers and productivity in italian manufacturing
  firms.
\newblock {\em Journal of productivity analysis}, 41(3):419--434, 2014.

\bibitem{raymond2015dynamic}
Wladimir Raymond, Jacques Mairesse, Pierre Mohnen, and Franz Palm.
\newblock Dynamic models of r \& d, innovation and productivity: Panel data
  evidence for dutch and french manufacturing.
\newblock {\em European Economic Review}, 78:285--306, 2015.

\bibitem{damijan2015learning}
Jo{\v{z}}e~P Damijan and {\v{C}}rt Kostevc.
\newblock Learning from trade through innovation.
\newblock {\em Oxford bulletin of economics and statistics}, 77(3):408--436,
  2015.

\bibitem{hall2003adoption}
Bronwyn~H Hall and Beethika Khan.
\newblock Adoption of new technology.
\newblock Technical report, National bureau of economic research, 2003.

\bibitem{gurbiel2002impact}
Roman Gurbiel.
\newblock Impact of innovation and technology transfer on economic growth: the
  central and eastern europe experience.
\newblock {\em Warshaw School of Economics}, 162:1--18, 2002.

\bibitem{surinach2011extension}
Jordi Suri{\~n}ach, Fabio Manca, Rosina Moreno, et~al.
\newblock Extension of the study on the diffusion of innovation in the internal
  market.
\newblock Technical report, Directorate General Economic and Financial Affairs
  (DG ECFIN), European~…, 2011.

\bibitem{moreno2014innovation}
Rosina Moreno~Serrano and Jordi Suri{\~n}ach~Caralt.
\newblock Innovation adoption and productivity growth: Evidence for europe
  (wp).
\newblock {\em AQR--Working Papers, 2014, AQR14/08}, 2014.

\bibitem{becheikh2006lessons}
Nizar Becheikh, Rejean Landry, and Nabil Amara.
\newblock Lessons from innovation empirical studies in the manufacturing
  sector: A systematic review of the literature from 1993--2003.
\newblock {\em Technovation}, 26(5-6):644--664, 2006.

\bibitem{singh2005collaborative}
Jasjit Singh.
\newblock Collaborative networks as determinants of knowledge diffusion
  patterns.
\newblock {\em Management science}, 51(5):756--770, 2005.

\bibitem{teece1992competition}
David~J Teece.
\newblock Competition, cooperation, and innovation: Organizational arrangements
  for regimes of rapid technological progress.
\newblock {\em Journal of economic behavior \& organization}, 18(1):1--25,
  1992.

\bibitem{tether2002co}
Bruce~S Tether.
\newblock Who co-operates for innovation, and why: an empirical analysis.
\newblock {\em Research policy}, 31(6):947--967, 2002.

\bibitem{brown1995product}
Shona~L Brown and Kathleen~M Eisenhardt.
\newblock Product development: Past research, present findings, and future
  directions.
\newblock {\em Academy of management review}, 20(2):343--378, 1995.

\bibitem{stock2002firm}
Gregory~N Stock, Noel~P Greis, and William~A Fischer.
\newblock Firm size and dynamic technological innovation.
\newblock {\em Technovation}, 22(9):537--549, 2002.

\bibitem{schumpeter1934theory}
JA~Schumpeter.
\newblock The theory of economic development. 7th edn (transl. opie r) harvard
  university press: Cambridge, 1934.

\bibitem{schumpeter1942socialism}
Joseph~Alois Schumpeter.
\newblock {\em Socialism, capitalism and democracy}.
\newblock Harper and Brothers, 1942.

\bibitem{tether1998small}
Bruce~S Tether.
\newblock Small and large firms: sources of unequal innovations?
\newblock {\em Research Policy}, 27(7):725--745, 1998.

\bibitem{acs1990innovation}
Zoltan~J Acs and David~B Audretsch.
\newblock {\em Innovation and small firms}.
\newblock Mit Press, 1990.

\bibitem{pavitt1987size}
Keith Pavitt, Michael Robson, and Joe Townsend.
\newblock The size distribution of innovating firms in the uk: 1945-1983.
\newblock {\em The Journal of Industrial Economics}, pages 297--316, 1987.

\bibitem{kleinknecht1993collecting}
Alfred Kleinknecht, Jeroen~ON Reijnen, and Wendy Smits.
\newblock Collecting literature-based innovation output indicators. the
  experience in the netherlands.
\newblock In {\em New concepts in innovation output measurement}, pages 42--84.
  Springer, 1993.

\bibitem{santarelli1996analyzing}
Enrico Santarelli and Roberta Piergiovanni.
\newblock Analyzing literature-based innovation output indicators: the italian
  experience.
\newblock {\em Research Policy}, 25(5):689--711, 1996.

\bibitem{cogan1993irish}
DJ~Cogan.
\newblock The irish experience with literature-based innovation output
  indicators.
\newblock In {\em New concepts in innovation output measurement}, pages
  113--137. Springer, 1993.

\bibitem{rogers2004networks}
Mark Rogers.
\newblock Networks, firm size and innovation.
\newblock {\em Small business economics}, 22(2):141--153, 2004.

\bibitem{revilla2012relation}
Antonio~J Revilla and Zulima Fern{\'a}ndez.
\newblock The relation between firm size and r\&d productivity in different
  technological regimes.
\newblock {\em Technovation}, 32(11):609--623, 2012.

\bibitem{minguela2014cooperation}
Beatriz Minguela-Rata, Jose Fern{\'a}ndez-Men{\'e}ndez, and Marta
  Fossas-Olalla.
\newblock Cooperation with suppliers, firm size and product innovation.
\newblock {\em Industrial Management \& Data Systems}, 2014.

\bibitem{pavitt1985patent}
Keith Pavitt.
\newblock Patent statistics as indicators of innovative activities:
  possibilities and problems.
\newblock {\em Scientometrics}, 7(1-2):77--99, 1985.

\bibitem{griliches1998patent}
Zvi Griliches.
\newblock Patent statistics as economic indicators: a survey.
\newblock In {\em R\&D and productivity: the econometric evidence}, pages
  287--343. University of Chicago Press, 1998.

\bibitem{pilkington2004technology}
Alan Pilkington.
\newblock Technology portfolio alignment as an indicator of commercialisation:
  an investigation of fuel cell patenting.
\newblock {\em Technovation}, 24(10):761--771, 2004.

\bibitem{fleming2007small}
Lee Fleming, Charles King~III, and Adam~I Juda.
\newblock Small worlds and regional innovation.
\newblock {\em Organization Science}, 18(6):938--954, 2007.

\bibitem{schilling2007interfirm}
Melissa~A Schilling and Corey~C Phelps.
\newblock Interfirm collaboration networks: The impact of large-scale network
  structure on firm innovation.
\newblock {\em Management science}, 53(7):1113--1126, 2007.

\bibitem{ma2008patent}
Zhenzhong Ma and Yender Lee.
\newblock Patent application and technological collaboration in inventive
  activities: 1980--2005.
\newblock {\em Technovation}, 28(6):379--390, 2008.

\bibitem{zhang2017network}
Gupeng Zhang, Hongbo Duan, and Jianghua Zhou.
\newblock Network stability, connectivity and innovation output.
\newblock {\em Technological Forecasting and Social Change}, 114:339--349,
  2017.

\bibitem{newman2001structure}
Mark~EJ Newman.
\newblock The structure of scientific collaboration networks.
\newblock {\em Proceedings of the national academy of sciences},
  98(2):404--409, 2001.

\bibitem{newman2002random}
Mark~EJ Newman, Duncan~J Watts, and Steven~H Strogatz.
\newblock Random graph models of social networks.
\newblock {\em Proceedings of the National Academy of Sciences}, 99(suppl
  1):2566--2572, 2002.

\bibitem{barabasi2002evolution}
Albert-Laszlo Barab{\^a}si, Hawoong Jeong, Zoltan N{\'e}da, Erzsebet Ravasz,
  Andras Schubert, and Tamas Vicsek.
\newblock Evolution of the social network of scientific collaborations.
\newblock {\em Physica A: Statistical mechanics and its applications},
  311(3-4):590--614, 2002.

\bibitem{albert2002statistical}
R{\'e}ka Albert and Albert-L{\'a}szl{\'o} Barab{\'a}si.
\newblock Statistical mechanics of complex networks.
\newblock {\em Reviews of modern physics}, 74(1):47, 2002.

\bibitem{bettencourt2009scientific}
Lu{\'\i}s~MA Bettencourt, David~I Kaiser, and Jasleen Kaur.
\newblock Scientific discovery and topological transitions in collaboration
  networks.
\newblock {\em Journal of Informetrics}, 3(3):210--221, 2009.

\bibitem{liu2015evolutionary}
Liang Liu, Chuanfeng Han, and Weisheng Xu.
\newblock Evolutionary analysis of the collaboration networks within national
  quality award projects of china.
\newblock {\em International journal of project management}, 33(3):599--609,
  2015.

\bibitem{liu2015structure}
Peng Liu and Haoxiang Xia.
\newblock Structure and evolution of co-authorship network in an
  interdisciplinary research field.
\newblock {\em Scientometrics}, 103(1):101--134, 2015.

\bibitem{wipoIPC}
International patent classification (ipc).
\newblock \url{https://www.wipo.int/classifications/ipc/en}.

\bibitem{choi2015predictive}
Jaehyun Choi, Dongsik Jang, Sunghae Jun, and Sangsung Park.
\newblock A predictive model of technology transfer using patent analysis.
\newblock {\em Sustainability}, 7(12):16175--16195, 2015.

\bibitem{jun2014small}
Sunghae Jun and Seung-Joo Lee.
\newblock A small world network for technological relationship in patent
  analysis.
\newblock In {\em Soft Computing in Big Data Processing}, pages 91--99.
  Springer, 2014.

\bibitem{park2015network}
Sangsung Park, Seung-Joo Lee, and Sunghae Jun.
\newblock A network analysis model for selecting sustainable technology.
\newblock {\em Sustainability}, 7(10):13126--13141, 2015.

\bibitem{guan2012patent}
Jiancheng Guan and Zifeng Chen.
\newblock Patent collaboration and international knowledge flow.
\newblock {\em Information Processing \& Management}, 48(1):170--181, 2012.

\bibitem{lissoni2010ape}
Francesco Lissoni, M~Coffano, A~Maurino, M~Pezzoni, and G~Tarasconi.
\newblock Ape-inv’s “name game” algorithm challenge: A guideline for
  benchmark data analysis and reporting.”.
\newblock Technical report, Technical Report, Academic Patenting in
  Europe-APE-INV, 2010.

\bibitem{maraut2008oecd}
St{\'e}phane Maraut, H{\'e}l{\`e}ne Dernis, Colin Webb, Vincenzo Spiezia, and
  Dominique Guellec.
\newblock The oecd regpat database: a presentation.
\newblock {\em OECD Science, Technology and Industry Working Papers}, 2008.
\newblock doi: \url{10.1787/241437144144}.

\bibitem{stauffer2018introduction}
Dietrich Stauffer and Ammon Aharony.
\newblock {\em Introduction to percolation theory}.
\newblock CRC press, 2018.

\bibitem{bunde2012fractals}
Armin Bunde and Shlomo Havlin.
\newblock {\em Fractals and disordered systems}.
\newblock Springer Science \& Business Media, 2012.

\bibitem{grimmett1999percolation}
Geoffrey Grimmett.
\newblock What is percolation?
\newblock In {\em Percolation}, pages 1--31. Springer, 1999.

\bibitem{ben2000diffusion}
Daniel Ben-Avraham and Shlomo Havlin.
\newblock {\em Diffusion and reactions in fractals and disordered systems}.
\newblock Cambridge university press, 2000.

\bibitem{dorogovtsev2002evolution}
Sergey~N Dorogovtsev and Jose~FF Mendes.
\newblock Evolution of networks.
\newblock {\em Advances in physics}, 51(4):1079--1187, 2002.

\bibitem{cohen2010complex}
Reuven Cohen and Shlomo Havlin.
\newblock {\em Complex networks: structure, robustness and function}.
\newblock Cambridge university press, 2010.

\bibitem{newman2018networks}
Mark Newman.
\newblock {\em Networks: an introduction}.
\newblock Oxford university press, 2018.

\bibitem{li2015network}
Daqing Li, Qiong Zhang, Enrico Zio, Shlomo Havlin, and Rui Kang.
\newblock Network reliability analysis based on percolation theory.
\newblock {\em Reliability Engineering \& System Safety}, 142:556--562, 2015.

\bibitem{karrer2014percolation}
Brian Karrer, Mark~EJ Newman, and Lenka Zdeborov{\'a}.
\newblock Percolation on sparse networks.
\newblock {\em Physical review letters}, 113(20):208702, 2014.

\bibitem{kawamoto2015precise}
Hirokazu Kawamoto, Hideki Takayasu, Henrik~Jeldtoft Jensen, and Misako
  Takayasu.
\newblock Precise calculation of a bond percolation transition and survival
  rates of nodes in a complex network.
\newblock {\em PloS one}, 10(4):e0119979, 2015.

\bibitem{newman2001scientific}
Mark~EJ Newman.
\newblock Scientific collaboration networks. i. network construction and
  fundamental results.
\newblock {\em Physical review E}, 64(1):016131, 2001.

\bibitem{tomassini2007empirical}
Marco Tomassini and Leslie Luthi.
\newblock Empirical analysis of the evolution of a scientific collaboration
  network.
\newblock {\em Physica A: Statistical Mechanics and its Applications},
  385(2):750--764, 2007.

\bibitem{do2013percolation}
Su~Do~Yi, Woo~Seong Jo, Beom~Jun Kim, and Seung-Woo Son.
\newblock Percolation properties of growing networks under an achlioptas
  process.
\newblock {\em EPL (Europhysics Letters)}, 103(2):26004, 2013.

\bibitem{li2015percolation}
Daqing Li, Bowen Fu, Yunpeng Wang, Guangquan Lu, Yehiel Berezin, H~Eugene
  Stanley, and Shlomo Havlin.
\newblock Percolation transition in dynamical traffic network with evolving
  critical bottlenecks.
\newblock {\em Proceedings of the National Academy of Sciences},
  112(3):669--672, 2015.

\end{thebibliography}

\appendix
\section{The EPO patent data}\label{sec:epoData}

The theme of this study is to investigate the evolution of the collaborations resulting from - and in - any kind of innovation focusing mainly on the European region. To this end, we employed the data set of patent applications to the European Patent Office (EPO) in order to build and study the characteristics of the EPO patent applicants’ network through time.

The specific goals of the project involved gathering and processing certain pieces of information, such as: which applicants contributed to which patent, the geographical origin of the applicants, the patent’s filing date and the corresponding technological areas of the patent. This data was derived from a combination of the EPO-subsets of two distinct databases: a) the “OECD, REGPAT database, July 2014” and b) the “OECD, Triadic Patent Families database, July 2014”.

The OECD REGPAT is a rich database which comprises two data sets: the set of patent applications to the EPO and those filed under the Patent Cooperation Treaty (PCT). The Triadic database consists of patents filed at all three of the following patent-granting organizations: the EPO, the United States Patent and Trademark Office (USPTO) and the Japan Patent Office (JPO), by the same applicant, while referring to the same invention.

At first glance, the REGPAT database fits perfectly to the needs of this study, providing all the required information: a list of all the applicants of each patent, regional data for each applicant, the patent filing and priority dates and information about the technological areas of a patent, in the form of International Patent Classification (IPC) codes. However, the OECD REGPAT database had two major limitations that we needed to address.

The first one is that the applicants are not uniquely identified. Each new entry is assigned a surrogate key derived by the combination of three fields: name, address and country code, which are entered by the applicant. Two entries, corresponding to the same applicant, with the slightest difference in either one of these three fields - e.g. an extra comma - appear as two distinct applicants in the database~\cite{lissoni2010ape}. Therefore, as multiple keys are very often assigned to the same applicant, a network derived directly from this database would be very different than the real one. To find the unique applicants, we chose three fields of each entry (name, address and NUTS code) as similarity indicators. Thankfully, the address and NUTS code of an applicant of the REGPAT database are very reliable entries~\cite{maraut2008oecd}.

We first prepared the database by applying preliminary cleaning procedures, such as removing corrupted or unwanted characters and normalizing the text by replacing umlauts, accents, etc. Then, we separated the applicants into groups according to their country code. For each applicant entry in a group, we calculated the Levenshtein distance of the three chosen fields with every other applicant entry. Whenever a pair of applicants scored lower than a certain threshold in all three distances, the applicants were considered to be identical. By running the algorithm on a large number of random samples and carefully inspecting the results, we managed to determine a threshold value that ensures avoidance of false positives. This procedure resulted in a remarkable reduction ($\sim$26\%) in the number of unique applicants.

The second limitation of the REGPAT database is that only the year of the patent filing and priority dates are explicitly provided. This was a major drawback, as we wanted to perform a temporal analysis, which requires a fine date granularity. We managed to successfully address this limitation by matching the EPO REGPAT patent set with the EPO TRIADIC patent set, which provides the full filing and priority dates, and by taking advantage of the implicit time-stamp of the EPO application ID.

In particular, we implemented the following procedure: First, we matched the TRIADIC entries to the REGPAT entries, by using the EPO application ID and appended the first priority date and the first EPO filing date of the TRIADIC entries to the REGPAT ones. Clearly, the TRIADIC EPO patent set is a subset of the REGPAT one, therefore after this step there were REGPAT entries still left unmatched. We determined the date of the unmatched entries by utilizing the fact that the EPO application ID is a seven digit serial number that holds implicit information about the chronological order of the patents and that the first EPO filing date is more consistent with the date mentioned in the EPO application ID than the first priority date.

Under these assumptions, we sorted all the entries – matched and unmatched – by two fields: the 11,546 matched first-EPO-filing dates and their EPO application ID. The result was a database of chronologically sorted patents, with a mix of matched entries and some sporadic blocks of a few unmatched entries. We inspected the distance between the consecutive sorted matched dates and noticed that the vast majority of them were only one day apart. This was clearly a very satisfactory level of granularity, therefore it was reasonable to consider the blocks of unmatched entries as having being filed on the same day of the last preceding matched entry. The network was built using this order.

\section{Confirmation of the percolation threshold}\label{sec:percConfirm}

Percolation theory models the behaviour of systems with metric structure, such as lattices, in which their sites/bonds are occupied with a probability $p$~\cite{stauffer2018introduction,bunde2012fractals,grimmett1999percolation,ben2000diffusion}. More specifically, it studies the emergence of a Giant Connected Component (GCC) – which is of the order of the system’s size – with respect of p. The GCC emerges at the percolation threshold of a critical probability $p_c$; below that threshold it does not exist.

A similar concept is also applicable in equilibrium (static) networks~\cite{newman2001structure,barabasi2002evolution,dorogovtsev2002evolution,cohen2010complex,newman2018networks}. As in the case of lattices, a connected component that contains a constant fraction of the network’s nodes can also exist in such networks. This connected component is called the giant component. Site percolation in networks can be described as the random removal of a network’s nodes and corresponding edges, with probability p. As more nodes are being removed, the network disintegrates into a sea of finite clusters. The phase transition between the two phases, the one in which a giant component exists, and the one in which the network is entirely fragmented and non-functional, is defined by the percolation threshold $p_c$.

For random, \textit{static} networks the percolation threshold can be determined by examining the size of the second largest cluster of the network~\cite{bunde2012fractals}. Specifically, as nodes are continuously removed from the network, the size of the second largest cluster varies, and it reaches a maximum at the percolation threshold, coinciding with the decomposition of the giant component and the complete fragmentation of the network into a sea of small-sized clusters. This criterion has been repeatedly used as a means to pinpoint the percolation threshold of real networks, including scale-free ones, such as the Internet~\cite{li2015network,karrer2014percolation,kawamoto2015precise}.

In \textit{growing} (evolving) networks, percolation is regarded as the emergence of the giant component from the coalescence of dominant clusters that grew out of an initial sea of small island clusters. In the evolving network of patents that we studied, there are always two dominant clusters, the union of which marks the birth of the giant component. We regard the patent that joins these two major clusters as the “critical” or “red-bond” patent. The giant component, once formed, quickly outgrows any other cluster in the system (Fig.~\ref{fig:LCgrowth} and Fig.~\ref{fig:LCgrowthDetail}). In order to pinpoint the red-bond patent, we follow the evolution of two largest clusters at any time-step, throughout the timeline, as in~\cite{bettencourt2009scientific,liu2015evolutionary,liu2015structure,newman2001scientific,tomassini2007empirical}.

\begin{figure}[ht]
    \centering
    \includegraphics[width=0.9\textwidth]{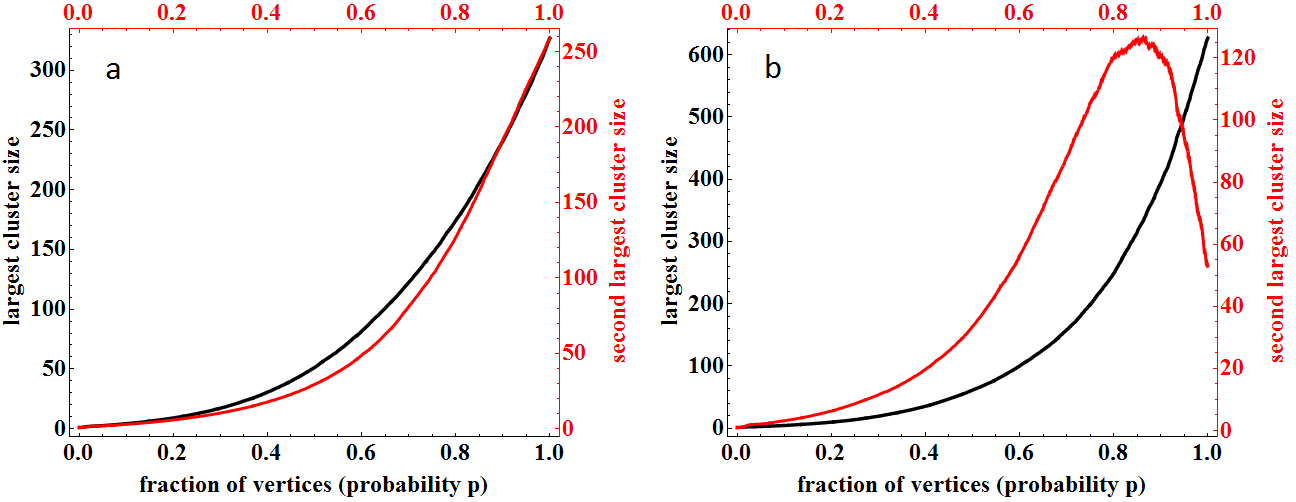}
    \caption[Damage report!]{Size of largest and second largest components, corresponding to snapshots of the growing network at points (a) below and (b) at and above the addition of the red-bond bond patent. \label{fig:confirmPercThreshold}}
\end{figure}

To confirm the validity of the red-bonds thus pinpointed, we employed the following criterion. The timeline is divided into two regimes: one starting from the beginning of the timeline up until right before the critical point and one from the critical point up to the end of the timeline. We take snapshots of the evolving network, distributed in both areas, including one on and exactly before the criticality. These snapshots are static (equilibrium) networks, on which we can apply the method of the second largest cluster size maximum, to determine the percolation threshold. Clearly, there should be a percolation threshold, $p_c$, denoted by the second largest cluster size maximum, for all snapshots in the second period, however, the second largest cluster should not have any maximum, in all snapshots of the first period. The results validated all the red-bonds previously found. In all snapshots below the critical point, i.e. before the addition of the red-bond patent, both clusters grow continuously with p, Fig.~\ref{fig:confirmPercThreshold}a. In contrast, the size of the second largest cluster increases until it reaches a maximum and then falls, in all snapshots at and above the red-bond patent Fig.~\ref{fig:confirmPercThreshold}b. This method was also used in other studies involving growing networks~\cite{do2013percolation,li2015percolation}. 

\section{Supplementary Figures}\label{sec:SupplFigs}

\begin{figure}[ht]
    \centering
    \includegraphics[width=0.55\textwidth]{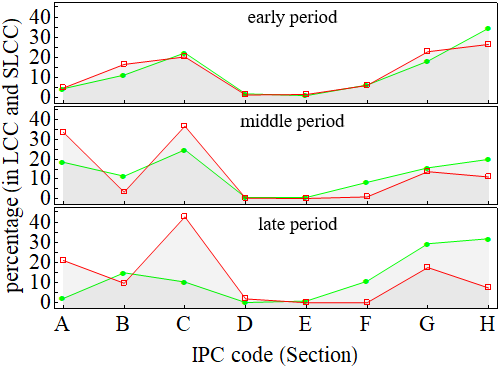}
    \caption[Damage report!]{Percentages of IPC codes in the two major groups of applicants (LCC and SLCC) at the adjacent-pre-percolation state, in three representative time-windows, one for each period (early, middle and late). \label{fig:sectionsInTime}}
\end{figure}

\begin{figure}[ht]
    \centering
    \includegraphics[width=0.55\textwidth]{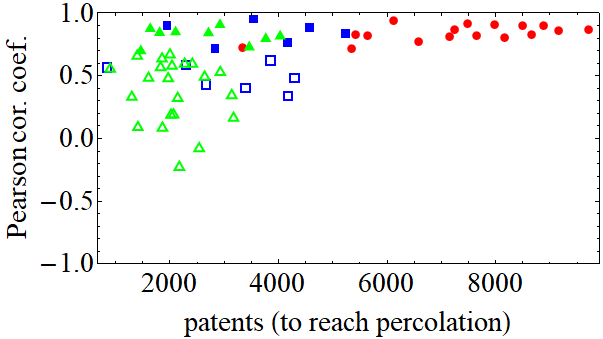}
    \caption[Damage report!]{Pearson correlation coefficient of IPC code percentages of largest and second largest connected component (LCC and SLCC) at their adjacent-pre-percolation state vs. number of patents required to reach percolation. Red, blue and green markers represent early, middle and late time-windows, respectively. Solid markers represent p-values that reject the null hypothesis (the two sets of values are independent) whereas the opposite holds for hollow markers. \label{fig:pearsonVsPatents}}
\end{figure}

\begin{figure}[ht]
    \centering
    \includegraphics[width=0.8\textwidth]{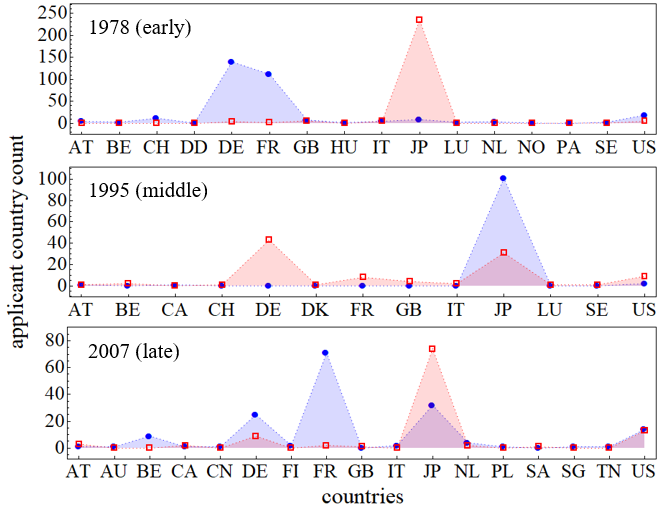}
    \caption[Damage report!]{Applicants' countries count in the two groups, i.e. the largest and second largest connected component (LCC and SLCC), immediately before the addition of the critical bond and the formation of the giant component (adjacent-pre-percolation state). One time-window for each of the three periods (early, middle and late). \label{fig:applicantCountryCountEML}}
\end{figure}

\begin{figure}[ht]
    \centering
    \begin{minipage}{0.49\textwidth}
        \centering
        \includegraphics[width=0.99\textwidth]{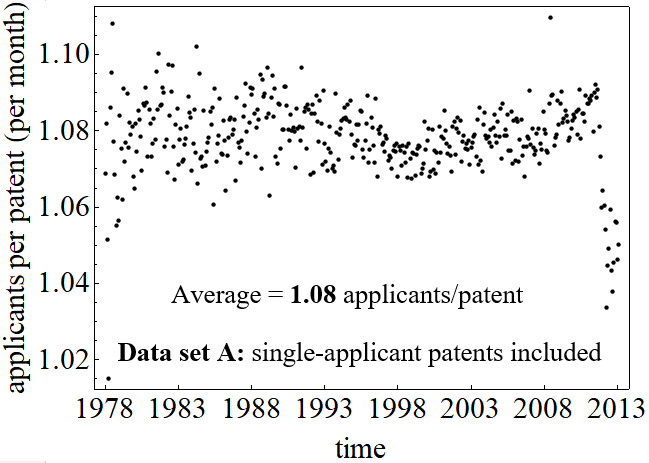} 
        \caption[Damage report!]{Number of applicants over submitted patents, averaged over each month, through time. An applicant 
can participate in multiple patents throughout the timeline, therefore the applicants are not unique. Data set A: all patents submitted in the 35 years of the timeline, including the ones with just one applicant. The latter result in isolated (island) nodes or self-loops in the network. Average: 1.08 applicants per patent.\label{fig:averageApplicantPerPatentDS_A}}
    \end{minipage}\hfill
    \begin{minipage}{0.49\textwidth}
        \centering
        \includegraphics[width=0.99\textwidth]{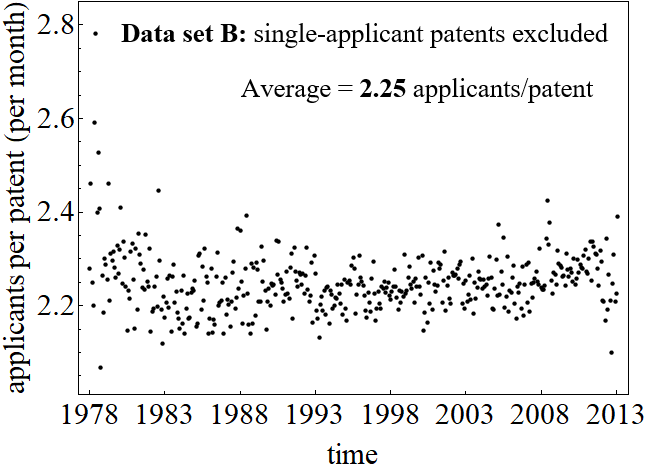} 
        \caption[Damage report!]{Number of applicants over submitted patents, averaged over each month, through time. An applicant 
can participate in multiple patents throughout the timeline, therefore the applicants are not unique. Data set B: all patents submitted in the 35 years of the timeline, excluding the ones with just one applicant. The latter result in isolated (island) nodes or self-loops in the network. Average: 2.25 applicants per patent.\label{fig:averageApplicantPerPatentDS_B}}
    \end{minipage}
\end{figure}

\begin{figure}[ht]
    \centering
    \includegraphics[width=0.6\textwidth]{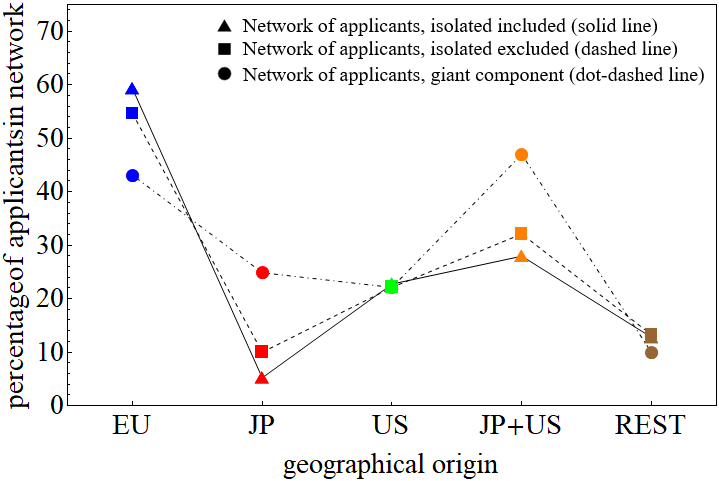}
    \caption[Damage report!]{Geographical breakdown of the (unique) applicants in the network resulting from the time-aggregated data, spanning all available timeline. (\textit{Triangles, solid line}: the network of applicants, isolated nodes included, \textit{Squares, dashed line}: the network of applicants, isolated nodes excluded, \textit{Disks, dot-dashed line}: the GC of the network) \label{fig:applicantsInNetGeogWL2}}
\end{figure}







\end{document}